\documentclass[published]{JHEP3} 

\JHEP{00(2010)000}

\usepackage{epsfig,multicol,bbm}

\newcommand\fverb{\setbox\fverbbox=\hbox\bgroup\verb}
\newcommand\fverbdo{\egroup\medskip\noindent%
			\fbox{\unhbox\fverbbox}\ }
\newcommand\fverbit{\egroup\item[\fbox{\unhbox\fverbbox}]}
\newbox\fverbbox

\title{A Scaling Relation of the Evolving Tidal Fields in a $\Lambda$CDM 
Cosmology}

\author{Jounghun Lee\\
Astronomy Program, FPRD, Department of Physics and Astronomy, Seoul 
National University, Seoul 151-747, Korea\\
E-mail: \email{jounghun@astro.snu.ac.kr}}

\author{Volker Springel\\
Max-Planck Institut fuer Astrophysik,
Karl-Schwarzschild Strasse 1, D-85748 Garching, Germany\\
E-mail: \email{volker@mpa-garching.mpg.de}}

\received{\today}               
\accepted{\today}               

\abstract{We report the finding of a scaling relation among the
 cosmic-web anisotropy parameter $A$, the linear density rms
 fluctuation $\sigma (r)$ and the linear growth factor $D(z)$. Using
 the tidal field derived from the Millennium Simulation on $512^{3}$
 grids at $z=0,\ 2,\ 5$ and $127$, we calculate the largest
 eigenvalues $\lambda$ of the local tidal tensor at each grid
 resolution and measure its distance-averaged two-point correlation
 function, $\xi_{\lambda}$, as a function of the cosines of polar
 angles $\cos\theta$ in the local principal axis frame.  We show
 that $\xi_{\lambda}$ is quite anisotropic, increasing toward the
 directions of minimal matter compression, and that the anisotropy of
 $\xi_{\lambda}$ increases as the redshift $z$ decreases and as the
 upper distance cutoff $r_{c}$ decreases.  Fitting the numerical
 results to an analytic fitting model
 $\xi_{\lambda}(\cos\theta)\propto (1+A\cos^{n}\theta)^{-1}$, it is
 found that the best fit value of $A$, dubbed the {\it cosmic-web
 anisotropy parameter}, varies systematically with $\sigma(r_{c})$ and
 $D(z)$, allowing  us to determine the simple empiral scaling
 relation $A(r_{c},z)=0.8\, D^{0.76}(z)\,\sigma (r_{c})$.}

\keywords{cosmic web, semi-analytic modeling}

\begin{document} 

\section{Introduction}

As confirmed by recent N-body
 simulations\cite{millennium05,simulation08}, the large-scale spatial
 distribution of cold dark matter exhibits an anisotropic web-like
 pattern, which is often dubbed the {\it cosmic web}.  The cosmic web
 of the dark matter distribution found in N-body simulations is also
 consistent with the observed large-scale filamentary distribution of
 galaxies in the real universe \cite{2df,sdss}. According to the
 standard model of cosmic structure formation \cite{bond-etal96}, the
 cosmic web originates in primordial density perturbations that are
 sharpened by gravitational tidal fields. The web becomes more
 anisotropic as the tidal fields become stronger due to the nonlinear
 processes during the gravitational evolution.
 
Various statistical tools have so far been proposed to describe the
 geometric properties of the cosmic web. For example, the conventional $N$-point
 statistics has been used to calculate the anisotropic spatial
 distribution of dark halos \cite{bon-mye96}. The Minkowski
 functionals have been employed to determine the morphological
 properties of the large-scale structures embedded in the cosmic web
 \cite{sch-etal99}. The $N$-dimensional skeleton approach has been
 found to be efficient in tracing the evolution of the cosmic web
 \cite{sou-etal08,sou-etal09}.  The Multiscale Morphology Filter
 method has been applied to the large-scale galaxy distribution to
 identify the anisotropic structures of the cosmic web
 \cite{ara-etal07}. A method utilizing the concept of the {\it Local
 Dimension} has been suggested for the local quantification of the
 shapes of the galaxy neighborhood in the cosmic web \cite{SB09}.  The
 ellipticity-ellipticity correlations of dark halos have been used to
 quantify the large-scale anisotropy of the cosmic web
 \cite{lee-etal08}. Finally, tessellation techniques have been suggested to
 trace the evolution of the geometric structures in the universe
 \cite{sha09,sha-etal09}.

Very recently, it has been pointed out by Lee, Hahn and Porciani 
\cite{lee-etal09}(LHP09, hereafter) that, since the cosmic web is 
produced by the anisotropic compression of matter along the principal axes 
of the large-scale tidal fields, its anisotropic nature may be best quantified 
in the system of the principal axes of the tidal fields. This led them to
suggest the anisotropic two-point correlations of the nonlinear traceless 
tidal field expressed in the principal axis frame, $\xi_{\lambda}({\bf r})$,  
as a new statistical tool for the description of the cosmic web phenomenon.

Analyzing numerical data from high-resolution N-body simulations, they have 
determined $\xi_{\lambda}({\bf r})$ and found that it has a much larger 
correlation length ($\sim 20\, h^{-1}$Mpc) and 
a higher degree of anisotropy than the density field itself. Integrating 
$\xi_{\lambda}({\bf r})$ over distance $r$ from $0$ to a certain cut-off scale 
$r_{c}$, and expressing it as a function of the angle between the major 
principal axes and the separation vectors, they noted that the nonlinear 
traceless tidal field has a much higher degree of anisotropy than the nonlinear 
density field. Interestingly, the results of LHC09 imply that the correlations of the 
nonlinear traceless tidal field may have a link to the initial conditions.

Yet, the analysis of LHP09 was restricted to the present epoch and
 their results were obtained by setting the distance cutoff scale
 $r_{c}$ to the correlation length of the nonlinear tidal field,
 $r_{c}=20\, h^{-1}$Mpc.  In this paper, our goal is to explore how
 the two-point correlations of the nonlinear traceless tidal field
 expressed in the principal axis system vary with redshift and
 distance scale by analyzing numerical data from high-resolution
 cosmological simulations. Throughout this paper, we assume a flat
 $\Lambda$CDM cosmology.

\section{Physical Analysis}

\subsection{Construction of the Tidal Fields}

For the construction of the tidal shear fields $T_{ij}({\bf x})$, we
 use the density contrast fields $\delta({\bf x})$ constructed on
 $512^{3}$ pixels from the Millennium Simulation
 \cite{millennium05} by means of the count-in-cell method, at four
 different redshifts $z=0,\ 2,\ 5$ and $127$.  The Millennium run
 \cite{millennium05} followed the evolution of the
 trajectories of $10^{10}$ dark matter particles, each of which has
 mass $8.6\times 10^{8}\, h^{-1}M_{\odot}$, in a periodic box of linear
 size $500\, h^{-1}$Mpc, using a $\Lambda$CDM cosmology with the 
 cosmological parameters
 $\Omega_{m}=0.25,\ \Omega_{\Lambda}=0.75,\ h=0.73,\ \sigma_{8}=0.9$
 and $n_{s}=1$.

The Fourier-transform of the density field, $\delta ({\bf k})$, is obtained 
through the Fast-Fourier-Transformation (FFT) method \cite{pre-etal92}. 
Then, the Fourier transform of the tidal shear field, $T_{ij}({\bf k})$, is 
calculated as $T_{ij}({\bf k})=k_{i}k_{j}\delta ({\bf k})/k^{2}$. The inverse 
Fourier-transformation of $T_{ij}({\bf k})$ yields the tidal shear field in 
real space, $T_{ij}({\bf x})$. The traceless tidal field is defined as
$\tilde{T}_{ij}({\bf x})\equiv T_{ij}({\bf x}) - \delta({\bf x})/3$.
At each pixel point we diagonalize $\tilde{T}_{ij}({\bf x})$ to find its three 
eigenvalues and the corresponding eigenvectors. The local principal axis frame 
has the three eigenvectors as basis vectors with polar axis in the direction 
of the eigenvector corresponding to the smallest eigenvalue. 
The eigenvector corresponding to the largest (smallest) eigenvalue is 
parallel to the direction of maximal (minimal) compression of local matter 
distribution. 

\subsection{Correlations of the Traceless Tidal Fields}

The two-point correlation function of the largest eigenvalue $\lambda$ of the local 
tidal tensor at redshift $z$ is defined as \cite{lee-etal09}
\begin{equation}
\label{eqn:xil}
\xi_{\lambda}({\bf r};z)=\langle\lambda ({\bf x};z)\cdot
\lambda ({\bf r}+{\bf x};z)\rangle.
\end{equation}
Let us express the separation vector ${\bf r}$ in terms of the spherical polar 
coordinates in the system of the principal axes of the local tidal tensor, 
${\bf r}=(r,\ \cos\theta,\ \phi)$, with the polar axis aligned with the 
direction of minimum matter compression, i.e., the direction of the 
eigenvector corresponding to the smallest eigenvalue of 
$\tilde{T}_{ij}({\bf x})$. If the $\lambda$-field was isotropic, then the 
two-point correlation $\xi_{\lambda}({\bf r})$ would depend only on the 
separation distance $r$, but not on the polar and azimuthal angles $\theta$ 
and $\phi$.  The degree of the anisotropy of the cosmic web can be quantified 
by measuring how strongly $\xi_{\lambda}$ changes with $\theta$ and $\phi$. 
\smallskip
\FIGURE{\epsfig{file=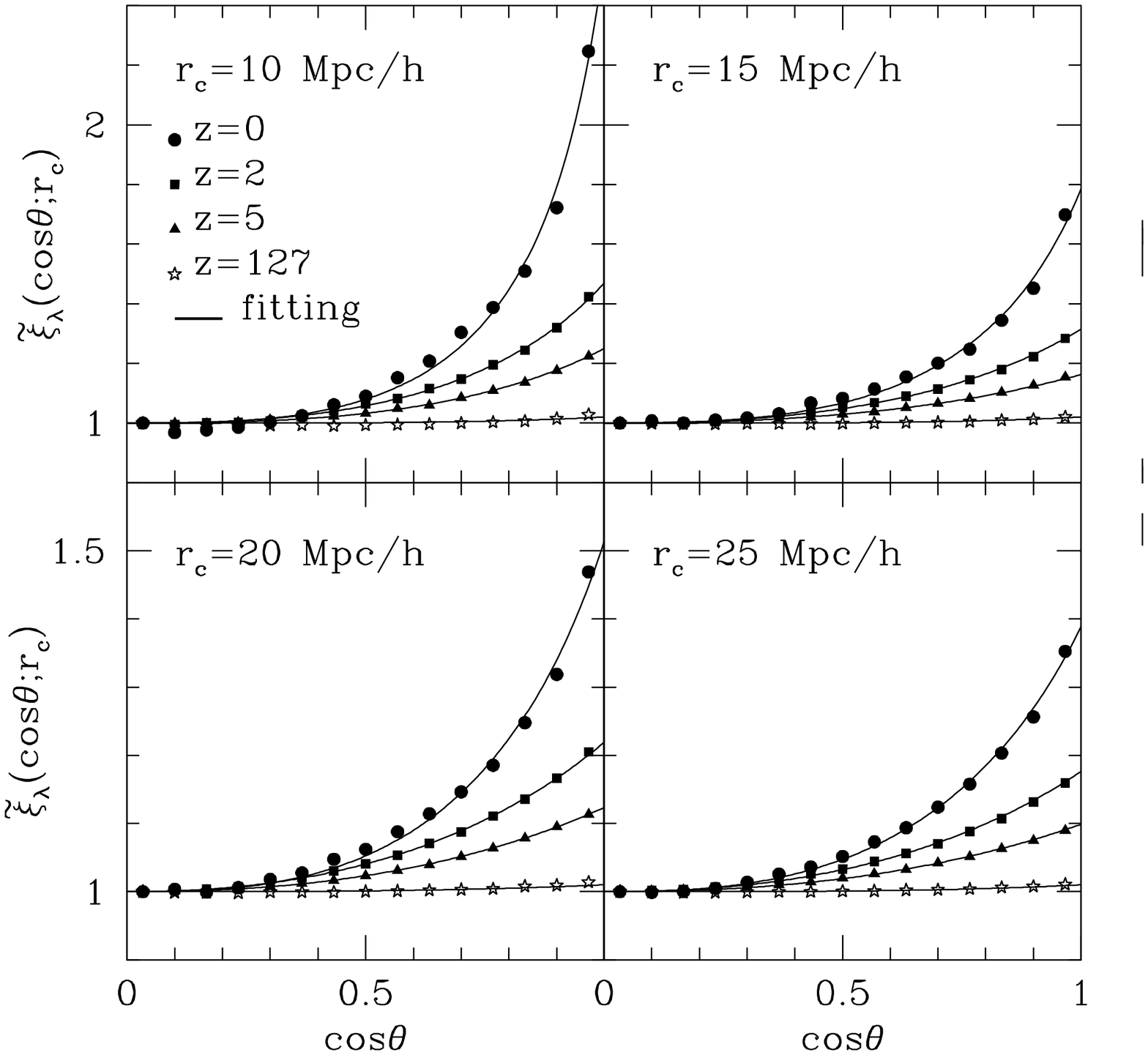,width=13cm} 
        \caption[Figure 1]
{Anisotropic two point correlations of the lowest eigenvalues of 
the traceless tidal field in the tidal principal axis frame averaged over 
the distance up to $r_{c}$ as a function of the cosine of the angles 
between the directions to the neighbor points and the eigenvectors 
corresponding to the lowest eigenvalues at four different redshifts 
($z=0,\ 2,\ 5$ and $127$ as dots, square dots, triangles and asterisks, 
respectively), for the four different values of the distance cut-off 
scale $r_{c}=10,\ 15,\ 20$ and $25\,h^{-1}$Mpc. 
In each panel, the thin solid lines represent the fitting models.}%
	\label{fig:ccor}}
As the dependence of $\xi_{\lambda}({\bf r})$ on the azimuthal angle $\phi$ 
has been found to be rather weak \cite{lee-etal09}, here we focus mainly on 
the polar-angle dependence of $\xi_{\lambda}({\bf r};z)$.

The anisotropic two-point correlation function of $\lambda$ as a function 
of the cosines of the polar angles can be obtained by averaging 
$\xi_{\lambda}({\bf r})$ over $r$ and $\phi$ as
\begin{equation}
\label{eqn:xil_rescaled}
\xi_{\lambda}(\cos\theta)\equiv\int^{r_{c}}_{0}{\rm
d}r\int^{2\pi}_{0}{\rm d}\phi
~\xi_{\lambda}(r,\cos\theta,\phi), 
\end{equation}
where $r_{c}$ denotes the upper distance cutoff. When we integrate 
$\xi_{\lambda}({\bf r})$ over $r$, we consider the distance cutoff scale 
$r_{c}$ greater than  $10\, h^{-1}$Mpc since $\xi_{\lambda}(\cos\theta)$ 
shows fluctuating unstable behavior at $r_{c}$ less than $10\, h^{-1}$Mpc. 
It is expected that the degree of the anisotropy of 
$\xi_{\lambda}(\cos\theta)$ depends on the distance cutoff $r_{c}$ and 
the redshift $z$, i.e., 
$\xi_{\lambda}(\cos\theta)=\xi_{\lambda}(\cos\theta;r_{c},z)$.

Using the data from the Millennium simulation, we numerically measure
$\xi_{\lambda}(\cos\theta;r_{c},z)$ for the cases of four different distance 
cutoff scales, $r_{c}=10,\ 15,\ 20$ and $25h^{-1}$Mpc, and at four different 
redshifts $z=0,\ 2,\ 5$ and $127$. For each pair of pixel points separated 
by ${\bf r}$, we first compute the product of the largest eigenvalues of the 
local traceless tidal tensors $\tilde{\bf T}$ and determine the polar and 
azimuthal angles of ${\bf r}$ in the local principal axes of the tidal tensor.
Then, we take the spatial average of it to calculate the correlation of the 
largest eigenvalue  $\xi_{\lambda}({\bf r};z)$. Finally, we determine the 
expression of the correlation as a function of the cosine of the polar 
angle by averaging $\xi_{\lambda}({\bf r};z)$ over the azimuthal angle 
$\phi$ and over distance $r$ from $0$ to $r_{c}$. 
For the detailed explanation on how to measure $\xi_{\lambda}(\cos\theta)$ 
from numerical data, we refer the readers to LHP09 \cite{lee-etal09}.

Figure \ref{fig:ccor} plots the rescaled anisotropic two-point correlation 
of the traceless tidal fields, 
$\tilde{\xi}_{\lambda}(\cos\theta)\equiv\xi_{\lambda}(\cos\theta)/
\xi_{\lambda}(0)$, for $r_{c}=10,\ 15,\ 20$ and $25\, h^{-1}$Mpc (in the 
top-left, top-right, bottom-left and bottom-right panels, respectively) 
at $z=0,\ 2,\ 5$ and $127$ (dots, squares, triangles and open stars, 
respectively). In each panel, the solid lines correspond to the fitting 
models described in section 3.1. As it can be seen, 
$\tilde{\xi}_{\lambda}(\cos\theta; r_{c},z)$ increases with 
$\cos\theta$ at all redshifts for all cases of $r_{c}$, indicating that 
the $\lambda$-field is anisotropic and more strongly correlated along the 
directions of minimum compression of dark matter in the local frame, which 
is consistent with the trend found by LHP09. 

The standard deviations on the numerical measurement of
 $\tilde{\xi}_{\lambda}(\cos\theta)$ are also calculated as
 statistical noises, which turn out to be as small as $\le 10^{-5}$
 for all cases. Although it would be desirable to estimate 
 jackknife error bars that include not only the statistical noise but
 also the cosmic variance, it is difficult to estimate them
 here since dividing the simulation box would destroy
 the periodicity of the box needed for our Fourier-based calculation
of the tidal field. Hence, we omit the errorbars in
 Fig.~\ref{fig:ccor} since the calculated statistical errors are anyway
 invisibly small.

\section{Results}

\subsection{Fitting Formula}

Noting that $\tilde{\xi}_{\lambda}(\cos\theta; r_{c})$ increases  
with $\cos\theta$, we employ the following fitting model for it 
\cite{lee-etal09}:
\begin{equation}
\label{eqn:xi_fit}
\tilde{\xi}_{\lambda}(\cos\theta,z;r_{c})=\frac{1}{1+A\cos^{n}\theta}, 
\end{equation}
 where $A$ and $n$ are two adjustable parameters. Fitting Equation
 (\ref{eqn:xi_fit}) to the numerical results with the help of the
 $\chi^{2}$-minimization method, we determine the best-fit values of
 $A$ and $n$ for each case of $r_{c}$ at each redshift.  For the
 calculation of $\chi^{2}$, we set the values of all errors associated
 with the numerical measurements to unity rather than to the
 corresponding standard deviations of $\tilde{\xi}_{\lambda}$
 \cite{BR96}, since the standard deviations are found to be extremely
 small in our case and do not account for cosmic variance, as
 discussed in section 2.2.  Table~\ref{tab:z0rc} lists the best-fit
 values obtained for $A$ and $n$ at $z=0$ for the four different values of
 $r_{c}$, while in Table~\ref{tab:rc20z} we provide the best-fit
 values of $A$ and $n$ at four different redshifts, setting $r_{c}$ to
 $20\, h^{-1}$Mpc.
\smallskip
\DOUBLETABLE
{\begin{tabular}{c|c|c}
	\hline	$r_{c}$ [$h^{-1}$Mpc] & $A$ & $n$ \\
	\hline $10$ & $0.61$  & $3.05$ \\
	$15$ & $0.43$  & $2.8$ \\
	$20$ & $0.34$  & $2.77$ \\
	$25$ & $0.28$  & $2.63$ \\
	\hline\end{tabular}}
{\begin{tabular}{c|c|c}
	\hline	redshift & $A$ & $n$ \\
	\hline $0$ & $0.34$  & $2.77$ \\
	$2$ & $0.18$  & $2.21$  \\
	$5$ & $0.11$  & $2.26$  \\
	$127$ & $0.01$  & $3.1$ \\
	\hline\end{tabular}}
{\label{tab:z0rc}Best-fit values of $A$ and $n$ at $z=0$ for 
the four distance cut-off scales $r_{c}$.}
{\label{tab:rc20z}Best-fit values of $A$ and $n$ with 
$r_{c}=20\, h^{-1}$Mpc at four different redshifts.}

Each panel of Fig.~\ref{fig:ccor} shows the fitting models of 
$\tilde{\xi}_{\lambda}$ with the best-fit values of $A$ and $n$ 
as thin solid lines. It can be seen that Eq.~(\ref{eqn:xi_fit}) indeed 
provides excellent fits to the numerical results for all values of $r_{c}$ 
at all redshifts. In all cases the minimum values of $\chi^{2}$ are formally 
found to be smaller than $10^{-3}$. 
The parameter $A$ is dubbed the {\it cosmic-web anisotropy 
parameter}, as it measures the strength of the dependence of 
$\tilde{\xi}_{\lambda}$ on $\cos\theta$. If the $\lambda$-field was
isotropic, $A$ would be zero. The more anisotropic the cosmic web is, 
the higher the value of this parameter $A$. Regarding the other fitting parameter  
$n$, it also shows a monotonic decrease with $r_{c}$, but no systematic 
change with redshift $z$.  Therefore, we focus mainly on the cosmic web anisotropy 
parameter $A$ from here on. 

It is worth comparing Eq.~(\ref{eqn:xi_fit}) with the fitting model given 
in LHP09, who defined the polar angle as the angle between ${\bf r}$ 
and the eigenvector corresponding to the {\it largest} eigenvalue. 
In this work, we define the polar angle as the angle between ${\bf r}$ and the 
eigenvector corresponding to the {\it smallest} eigenvalue.  The reason for 
using a different definition of the polar axis here is that 
$\xi_{\lambda}$ is found to exhibit more obvious anisotropic behavior when 
the polar axis is chosen to be parallel to the direction of the eigenvector 
corresponding to the smallest eigenvalue. In addition, we also vary
the power of $\cos\theta$ in the fitting model, while in LHP09 this
was kept fixed. 

\subsection{A Scaling Relation}

\smallskip
\FIGURE{\epsfig{file=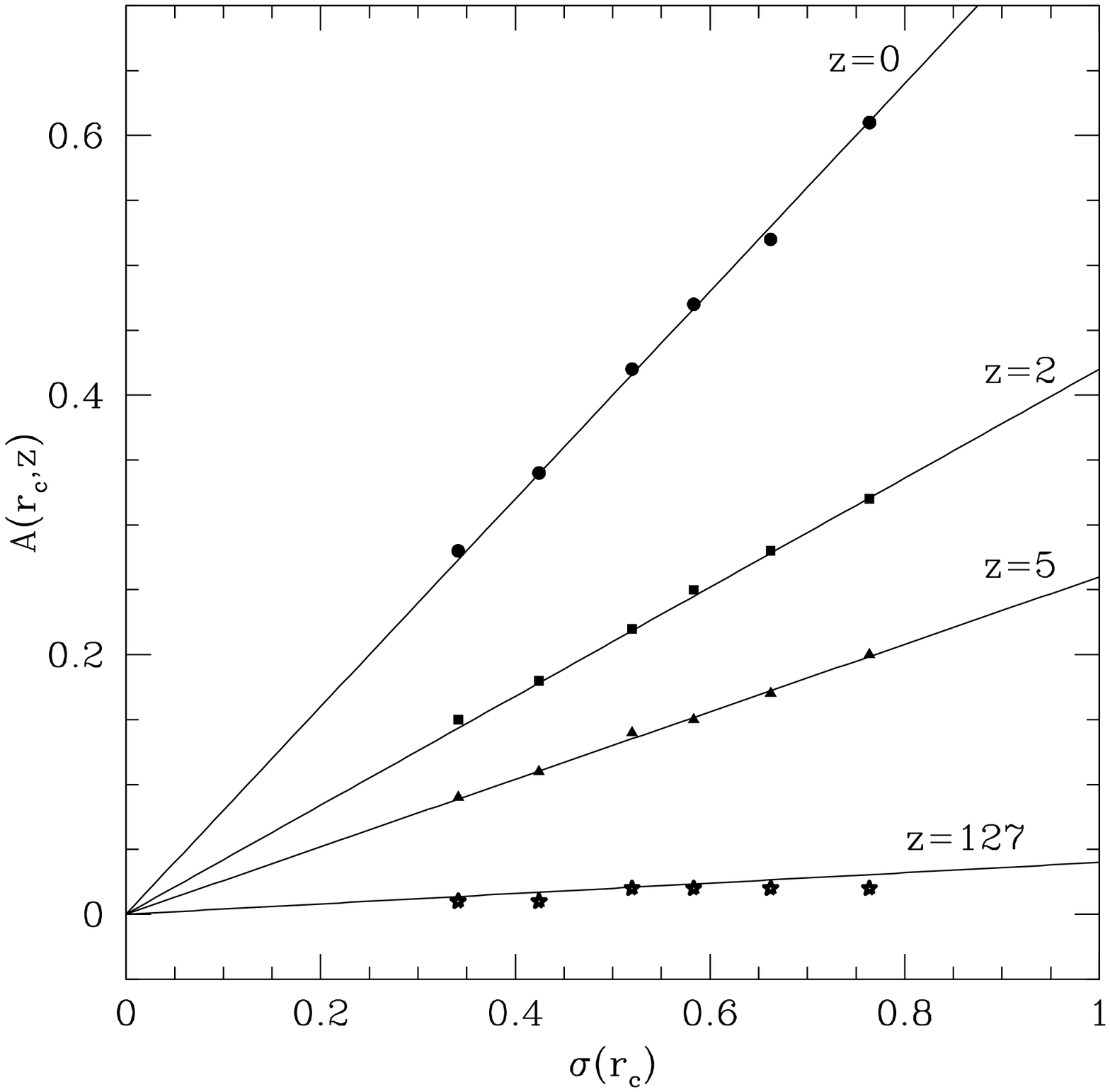,width=13cm} 
        \caption[Figure 2]
{Scaling relations between the cosmic web anisotropy parameter $A$ 
and the linear density rms density fluctuations on the distance cutoff 
scale $r_{c}$ at $z=0,\ 2,\ 5$ and $127$.}%
	\label{fig:fit_ccor}}
Noting a systematic trend in the variation of the cosmic-web anisotropy 
parameter $A$ with $r_{c}$ and $z$, we investigate how $A$ depends on the 
linear density rms fluctuation $\sigma (r_{c})$ smoothed on the distance 
cut-off scale $r_{c}$, and on the linear growth factor $D(z)$. Using various 
distance cut-off values equal to $r_{c}=10,\ 12,\ 14,\ 16,\ 20$ and $25\, h^{-1}$Mpc 
at each redshift, we measure numerically 
$\tilde{\xi}_{\lambda}(\cos\theta;r_{c})$ by using the data from the 
Millennium Simulation and by determining the best-fit value of $A$ 
through $\chi^{2}$ fitting of Eq.~(\ref{eqn:xi_fit}) to the numerical 
data points in each case. 

We also analytically compute the linear density rms fluctuations 
$\sigma (r_{c})$ smoothed on the same distance cut-off scale $r_{c}$ as  
\begin{equation}
\label{eqn:sig}
\sigma^{2}(r_{c})=\int_{-\infty}^{\infty}\Delta^{2}(k)
W^{2}(kr_{c})d\ln k,
\end{equation}
where $W(kr_{c})$ is the top-hat filter of scale radius $r_{c}$ and 
$\Delta^{2}(k)$ is the dimensionless linear matter power spectrum. 
The following analytic approximation for $\Delta^{2}(k)$ given 
by \cite{BBKS86} is used:
\begin{eqnarray}
\label{eqn:delk}
\Delta^{2}(k) &\propto& k^{n_s+3}
\left[\frac{\ln(1+2.34q)}{2.34q}\right]^{2}
[1+3.89q+(16.1q)^{2}+(5.46)^{3}+(6.71q)^{4}]^{-1/2},
\end{eqnarray}
where $q \equiv k/[\Omega_{m}h^{2}{\rm Mpc}^{-1}$] \cite{PD94} and  
$n_{s}$ is the spectral index of the primordial power spectrum. 
For the case that the key cosmological parameters ($\Omega_{m}$, $h$ and 
$\sigma_{8}$) are set at the values used by the Millennium run, this 
formula turns out to agree sufficiently well with the result from the 
CMBFAST code \cite{SZ96}. 

Figure \ref{fig:fit_ccor} plots the best-fit values of $A$ for the six
 different values of $r_{c}$ versus the linear rms density
 fluctuations $\sigma (r_{c})$ smoothed on the same scale, at the four
 different redshifts we considered. This plot reveals that the
 cosmic-web anisotropy parameter $A$ is directly proportional to
 $\sigma(r_{c})$. The cosmic web anisotropy parameter $A$ is shown to
 decrease monotonically with $z$, quantifying the evolution of the
 anisotropy of the traceless tidal fields.  Note that this result is
 consistent with the qualitative explanation of the cosmic web theory
 \cite{bond-etal96}.

It is interesting to see in Fig.~\ref{fig:fit_ccor} that $A[\sigma(r_{c})]$ 
at each redshift is well described by a straight line, and that the slope of 
the line decreases with $z$. Supposing $A\propto \sigma(r_{c})$, we 
investigate how the proportionality factor between $A$ and $\sigma(r_{c})$ 
(i.e., the slope of the straight line) changes with the linear growth factor 
$D(z)$. For the evaluation of the linear growth factor $D(z)$, we use 
the formula given in \cite{lahav-etal91}. 
\begin{equation}
\label{eqn:lcdmD}
D(z) \propto \frac{5}{2}\Omega_{m}
[\Omega_{m}(1+z)^{3}+\Omega_{\Lambda}]^{1/2}\int_{z}^{\infty}dz^{\prime}
\frac{1+z^{\prime}}{[\Omega_{m}(1+z^{\prime})^{3} + 
\Omega_{\Lambda}]^{3/2}}.
\end{equation}

We numerically calculate the ratio, $A/\sigma(r_{c})$, at each redshift.  
Fitting $A/\sigma(r_{c})$ to a power-law formula $\alpha D^{\beta}(z)$ 
and determining the best-fit value of $\alpha$ and 
$\beta$ at each redshift with the help of the $\chi^{2}$ statistics, 
we find that the ratio $A/\sigma(r_{c})$ varies as $0.8\,D^{0.76}(z)$. 
Fig.~\ref{fig:fit_ct} plots $A(r_{c},z)/\sigma (r_{c})$ 
versus $D(z)$. The numerical results of $A(r_{c},z)/\sigma (r_{c})$ at four 
redshifts are shown as circles while the fitting model $0.8\, D^{0.76}(z)$ is shown 
as a solid line. This result demonstrates that the numerically obtained ratio between 
$A$ and $\sigma(r_{c})$ is indeed well fitted by $0.8D^{0.76}(z)$. 
\smallskip
\FIGURE{\epsfig{file=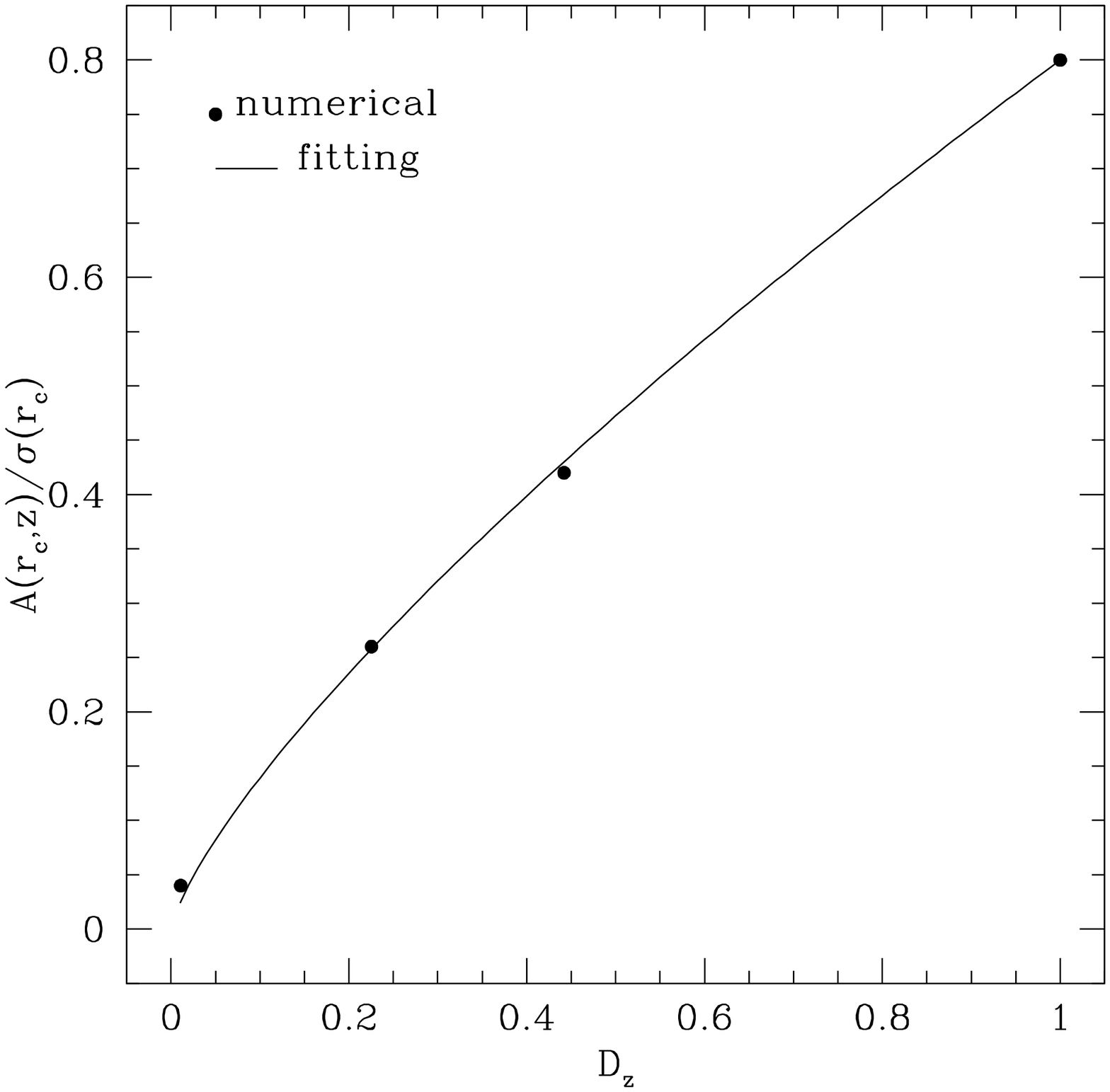,width=13cm} 
        \caption[Figure 3]
{Scaling relation between $A(r_{c},z)/\sigma(r_{c})$ and the 
linear growth factor $D_{z}$.}%
	\label{fig:fit_ct}}

Finally, we can combine these findings into the following scaling relation among cosmic web 
anisotropy parameter $A(r_{c},z)$, linear density rms fluctuation
$\sigma (r_{c})$ on the scale of $r_{c}$,  and linear growth factor $D(z)$:
\begin{equation}
\label{eqn:ar_scale}
A(r_{c},z)=0.8\, D^{0.76}(z)\sigma (r_{c}). \\
\end{equation}
This scaling relation quantifies how the cosmic web anisotropy parameter 
increases as the Universe evolves, revealing a link between the initial 
conditions and the anisotropic clustering of the nonlinear traceless tidal 
field expressed in the principal axis frame.

\section{Discussion}

To quantify the web-like pattern in the large-scale matter
 distribution, we have introduced the new concept of the cosmic web
 anisotropy parameter, which measures the degree of the anisotropy of
 the two-point correlations of the largest eigenvalues of the
 traceless tidal fields in the principal-axis frame. By analyzing the
 numerical data from the Millennium Simulation at different redshifts,
 we have empirically found a simple scaling relation among the cosmic
 web anisotropy parameter, the linear density rms fluctuations and the
 linear growth factor. Our results have allowed us to quantify how the
 anisotropy of the traceless tidal fields increases as the growth
 factor and the linear density rms fluctuations increase.  This new
 scaling relation thus provides a measure for the growth of anisotropy
 in the evolving cosmic web of the Universe.

The simple functional forms we found for the correlation and the
 scaling relation (eqs. [\ref{eqn:xi_fit}] and [\ref{eqn:ar_scale}])
 suggest that there might exist a simple explanation for these
 functional forms in linear perturbation theory. Yet, we have not been
 able to develop such a theory thus far, as the analytic treatment
 from first principles turns out to be extremely hard since we deal
 here with the {\it nonlinear} two-point correlations measured in the
 {\it principal} axis frame.

A crucial implication of our result is that the anisotropy of the
 cosmic web may be produced by three competing effects. The trace part
 of the tidal field and the cosmic expansion tend to make the matter
 distribution more isotropic, whereas the traceless part of the tidal
 field stretches the matter distribution and induces large-scale
 anisotropy. The competition among these three effects imprints the
 web-like pattern in the large-scale matter distribution, just as the
 competition between gravity and radiation pressure has left imprints
 in the form of acoustic oscillations in the temperature map of the
 cosmic microwave background radiation \cite{HS95}.

Since the linear growth factor $D(z)$ and the linear density rms
 fluctuations $\sigma(r_{c})$ are functions of the primary
 cosmological parameters such as the dark energy equation of state
 parameter $w$ and the density parameter $\Omega_{m}$, our results
 also hint that the cosmic web anisotropy parameter may be a new probe
 of cosmology. Given the scaling relation (eq.[\ref{eqn:ar_scale}]),
 we expect that if the cosmic web anisotropy parameter $A$ is measured
 for two different distance cut-off scales $r_{c}$ at the same
 redshift, it may be possible to put a constraint on the density
 parameter $\Omega_{m}$ by taking the ratio between the two values,
 removing the parameter degeneracy with the amplitude of the linear
 power spectrum $\sigma_{8}$. Similarly, by measuring the cosmic web
 anisotropy parameter at two different redshifts and taking the ratio
 between the two values, a constraint on the dark energy equation of
state parameter $w$ results.

The success of using the cosmic web anisotropy parameter as a new
 cosmological probe, however, is contingent upon a couple of future
 tests. First of all, it needs to be examined whether or not the same
 scaling relation also holds for different cosmologies since we have
 derived it assuming a $\Lambda$CDM cosmology with a specific set of
 cosmological parameters. Secondly, what is readily measurable in
 practice is not the tidal field of the dark matter distribution but
 that of the galaxy distribution.  It will hence be necessary to
 investigate how the bias between light and matter affects the scaling
 relation. We investigate this question in forthcoming work, and hope
 to report the results elsewhere in the near future.

\acknowledgments 

The Millennium Simulation data used in this 
work are available at http://www.mpa-garching.mpg.de/millennium. 
We thank an anonymous referee for helpful suggestions. 
J.L. is very grateful to S.D.M.White and the Max-Planck-Institute for 
Astrophysicsat in Garching for the warm hospitality where this research was 
initiated. J.L. acknowledges financial support from the Korea Science and 
Engineering Foundation (KOSEF) grant funded by the Korean Government 
(MOST, NO. R01-2007-000-10246-0).

\end{document}